\documentclass{aa} 

\newcommand{\Msunnom}{\hbox{$\mathcal{M}^{\rm N}_\odot$}}
\newcommand{\Rsunnom}{\hbox{$\mathcal{R}^{\rm N}_\odot$}}
\newcommand{\Teff}{\ensuremath{T_{\rm eff}}}                      
\newcommand{\kms}{\,km\,s$^{-1}$}                                 

\usepackage{graphicx}
\usepackage{txfonts}
\begin{document}

   \title{V456\,Cyg: An eclipsing binary\\with tidally perturbed g-mode pulsations}
  \titlerunning{Tidally perturbed pulsations in V456\,Cyg}

   \author{T. Van Reeth \inst{1}
          \and
           J. Southworth \inst{2}
           \and
        J. Van Beeck \inst{1}
           \and
        D. M. Bowman \inst{1}
          }

   \institute{Institute of Astronomy, KU Leuven, Celestijnenlaan 200D, B-3001 Leuven, Belgium\\
              \email{timothy.vanreeth@kuleuven.be}
             \and Astrophysics Group, Keele University, Staffordshire ST5 5BG, UK}

   \date{Received; accepted}
 
  \abstract
   {Many well-known bright stars have been observed by the ongoing Transiting Exoplanet Survey Satellite (TESS) space mission. For several of them, these new data reveal previously unobserved variability, such as tidally perturbed pulsations in close binary stars.}{Using newly detected gravity-mode (g-mode) pulsations in V456\,Cyg, we aim to determine the global stellar properties of this short-period eclipsing binary and evaluate the interaction between these pulsations and the tides.}{We model the binary orbit and determine the physical properties of the component stars using the TESS photometry and published spectroscopy. We then measure the pulsation frequencies from the residuals of the light curve fit using iterative prewhitening, and analyse them to determine the global asteroseismic stellar parameters. We evaluate the pulsation parameters as a function of the orbital phase.}{We find that the pulsations belong to the secondary component of V456\,Cyg and that this star likely has a uniform radial rotation profile, synchronous ($\nu_{\rm rot} = 1.113(14)\,\rm d^{-1}$) with the binary orbit ($\nu_{\rm orb} = 1.122091(8)\,\rm d^{-1}$). The observed g~modes are amplified by almost a factor three in the stellar hemisphere facing the primary. We present evidence that this is caused by tidal perturbation of the pulsations, with the mode coupling being strongly affected.}{V456\,Cyg is only the second object for which tidally perturbed high-order g-mode pulsations are identified, after $\pi^5$\,Ori. This opens up new opportunities for tidal g-mode asteroseismology, as it demonstrates another avenue in which g~modes and tides can interact with each other.}


   \maketitle
%
\section{Introduction}
\label{sec:intro}
In the study of stellar structure and evolution \citep[e.g.][]{Kippenhahn2012}, binary stars can both improve and complicate the analyses. By modelling the orbits using both photometric and spectroscopic observations, stellar properties such as mass and radius can be determined up to 1\% accuracy for the binary components \citep[e.g.][]{Torres2010}. However, binary interaction also affects stellar evolution, for example via tides \citep[e.g.][]{Ahuir2021,Zanazzi2021} and mass transfer \citep[e.g.][]{Shao2016,Vos2019}. These processes are not yet well understood, and given that many stars are part of multiple systems \citep[e.g.][]{Raghavan2010,Sana2012,MoeDiStefano2017}, this remains one of the most important questions in astronomy.

Asteroseismology, the study of stellar structure and evolution via the analysis of stellar pulsations \citep[e.g.][]{Aerts2010}, complements the study of binary stars \citep[e.g.][]{Beck2018,Johnston2019,Guo2021,Sekaran2021}. Gravity (g) modes, which have buoyancy as the dominant restoring force and are sensitive to the near-core radiative region in intermediate- to high-mass stars with convective cores such as $\gamma$\,Doradus \citep[$\gamma$\,Dor, with $1.4\,{\rm M}_\odot \lesssim M_* \lesssim 1.8\,{\rm M}_\odot$;][]{Kaye1999} and slowly pulsating B stars \citep[SPB, with $2.5\,{\rm M}_\odot \lesssim M_* \lesssim 8\,{\rm M}_\odot$;][]{Waelkens1991}, are especially valuable. They can be used to measure stellar age \citep[e.g.][]{Szewczuk2018,Mombarg2019,Wu2020,Michielsen2021,Pedersen2021}, rotation \citep[e.g.][]{Bouabid2013,VanReeth2016,Christophe2018,Li2020,Takata2020b,Takata2020a,Szewczuk2021}, near-core and envelope mixing \citep[e.g.][]{Szewczuk2018,Wu2019,Wu2020,Mombarg2020,Pedersen2021}, and internal magnetic fields \citep[e.g.][]{Buysschaert2018,Prat2019,Lecoanet2021}, thanks to high-quality space photometry from missions such as CoRoT \citep{Auvergne2009}, {\em Kepler} \citep{Borucki2010,Koch2010}, BRITE \citep{Weiss2014}, and TESS \citep{Ricker2015}. However, the pulsations themselves also affect stellar properties. For example, inertial and overstable convective modes in the convective core, which have the Coriolis force as the dominant restoring force, can couple with g~modes in the radiative envelope, leading to angular momentum transport between the core and the envelope \citep[e.g.][]{Ouazzani2020,Saio2021,Lee2021}.

In multiple systems, tides can also influence stellar pulsations. In eccentric binaries with a close periastron passage, dynamical tides can excite tidal oscillations, with frequencies at integer multiples of the  orbital frequency \citep[e.g.][]{Fuller2017,Hambleton2018,Cheng2020}. In other binaries, g-mode oscillations and tides couple non-linearly \citep[e.g.][]{Burkart2012,Burkart2014,Weinberg2013,Guo2017}, with sums of non-harmonic mode frequencies being equal to orbital harmonics. When tidal deformation of stars perturbs pulsation mode cavities, the pulsation frequencies split into $(\tilde{\ell}+1)$ multiplets \citep[e.g.][]{Reyniers2003b,Reyniers2003a,Balona2018}. These depend on the mode geometries ($\tilde{\ell},\tilde{m}$) with respect to the tidal axis, which connects the centres of mass of the binary components, where $\tilde{\ell}$ and $\tilde{m}$ indicate the spherical degree and azimuthal order, respectively. Stellar rotation further splits each of these frequencies into $(2\ell+1)$ multiplets, which are non-equidistant because of the Coriolis force \citep[e.g.][]{Bouabid2013}, in the observer's inertial reference frame. Here the mode geometry ($\ell,m$) is defined with respect to the rotational axis. \citet{Fuller2020} also included tidal pulsation coupling in this theoretical framework, demonstrating that this may amplify or trap pulsations in a hemisphere of the star, calling them tidally tilted pulsations. So far, tidal perturbation and tilting of pressure (p-)mode pulsations has been detected for about a dozen targets \citep[e.g.][]{Hambleton2013,Balona2018,Samadi2018,Bowman2019,Handler2020,Kurtz2020,Rappaport2021, Steindl2021,Southworth2020,Southworth2021, LeeWoo2021,Alicavus2022}, while tidally perturbed g~modes have only been found for one star \citep{Jerzykiewicz2020}.

In this work, we present the discovery of tidally perturbed g-mode pulsations in the eclipsing binary system V456\,Cyg. It was first identified as an eclipsing binary by \citet{Morgenroth1935} and classified as an Algol system. Following increasingly accurate measurements of the orbital period \citep{Savedoff1951,Whitney1959}, the detection of apsidal motion was reported \citep[e.g.][]{WoodForbes1963}. Over the following decades, V456\,Cyg was a target of several observational campaigns, both confirming \citep[e.g.][]{Flin1969,Mallama1980} and disproving \citep[e.g.][]{Hagedus1988} this detection. Using modern observations, \citet{Nelson2011} showed that the ephemerides are constant and determined that the orbit is quasi-circular, with an eccentricity of 0.0016\,(6). \citet{Bakis2014} confirmed these results after spectral disentangling of 12 low-resolution spectra ($R\sim5100$) from the T\"UB\.ITAK Faint Object Spectrograph and Camera (TFOSC) at the Russian and Turkish Telescope (RTT), located at the T\"UB\.ITAK National Observatory (TUG, Turkey). \citet{Bakis2014} determined the properties of the system, finding the primary to be a slightly metal-weak A-type star with a mass of $1.86\pm0.06$\,M$_\odot$ and the secondary to be an early F-type star with a mass of $1.58\pm0.05$\,M$_\odot$. The properties of the secondary put it either in or close to the $\gamma$\,Dor instability strip in the HR~diagram. Based on earlier measurements, \citet{Soydugan2006} had already classified V456\,Cyg as a candidate $\delta$\,Scuti pulsator. Although it is located within the instability strip, no pulsations were detected during following ground-based observational campaigns \citep{Dvorak2009,Liakos2009}. While a large fraction of stars in the $\delta$\,Scuti instability strip do not pulsate \citep{Murphy2019}, the cause and correlations with other stellar aspects, such as metallicity and binarity, are not yet known.

In the following sections, we outline the TESS observations (Sect.\,\ref{sec:tess-data}), measure the properties of the component stars (Sect.\,\ref{sec:binarity}), and perform an asteroseismic analysis of the TESS photometry (Sect.\,\ref{sec:asteroseismology}) of V456\,Cyg. We report the detection of g-mode pulsations (Sect.\,\ref{subsec:freq-analysis}), use them to measure the near-core rotation rate of the secondary (Sect.\,\ref{subsec:core-rot}), and provide evidence for their tidally perturbed nature (Sect.\,\ref{subsec:tidal-tilted}). Finally, we discuss our results and their implications for the tidal coupling of g-mode pulsations (Sect.\,\ref{sec:conclusions}).

\section{TESS photometry}
\label{sec:tess-data}

The TESS mission \citep{Ricker2015} has provided time series broad-band photometry of $\sim$85\,\% of the sky since July 2018. During the nominal mission, Full Frame Images (FFI) were taken at a 30-min cadence, yielding raw photometry for $\sim$10 million stars brighter than 13.5\,mag in the TESS passband \citep{Huang2020a,Huang2020b}. For over 200\,000 selected stars, ``postage stamp'' pixel data centred on the stars were also collected at a 2-min cadence \citep[e.g.][]{Stassun2018}. 

\begin{figure}
    \centering
    \includegraphics[width=\columnwidth]{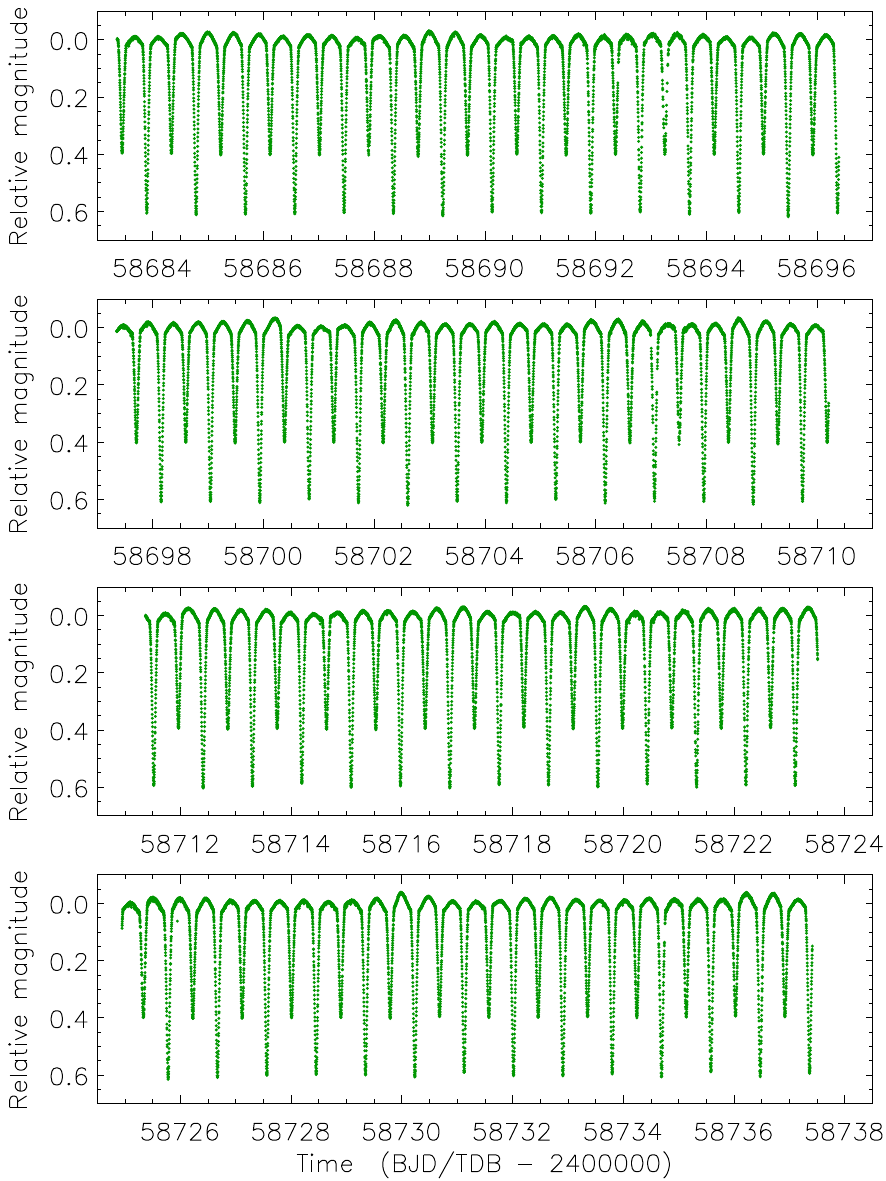}
    \caption{TESS light curves of V456\,Cyg from sectors 14 (top two panels) and 15 (bottom two panels). The data have been converted to relative magnitude and rectified to zero magnitude by subtraction of a straight-line fit to the data.}
    \label{fig:tess:lc}
\end{figure}

V456\,Cyg was observed by TESS during sectors 14 and 15 (2019 July 18 to 2019 Sep 11). These target pixel files (TPF) were reduced and light curves were extracted by the TESS Science Processing Operations Center \citep[SPOC;][]{Jenkins2016} using simple aperture photometry (SAP). In addition to these SAP light curves, the SPOC also provides Pre-search Data Conditioning (PDC) SAP light curves, optimised for the detection of exoplanet transits and from which instrumental variability was removed using cotrending base vectors. However, this reduction process can also affect intrinsic longer-period variability such as from binarity. Hence, we used the SAP light curve for the analysis of V456\,Cyg. The TESS data are shown in Fig.\,\ref{fig:tess:lc} after conversion to relative magnitude and linear detrending.

\section{Binary model}
\label{sec:binarity}

Our first analysis step was to model the effects of binarity in the TESS light curve for two reasons: to determine the physical properties of the stars and to obtain a light curve from which the effects of binarity had been removed. For this we used version 42 of the {\sc jktebop}\footnote{\texttt{http://www.astro.keele.ac.uk/jkt/codes/jktebop.html}} code \citep{Me++04mn2,Me13aa}. This code works with the fractional radii of the stars, $r_{\rm A}$ and $r_{\rm B}$, which are the true radii divided by the semi-major axis of the relative orbit. We fitted for their sum ($r_{\rm A}+r_{\rm B}$) and ratio ($k = \frac{r_{\rm B}}{r_{\rm A}}$), the orbital inclination ($i$), period ($P$), reference time of primary eclipse ($T_0$), the ratio of the central surface brightness values of the stars ($J$), and the third light ($L_3$). We refer to the hotter and more massive star as star~A and its companion as star~B.

The orbital eccentricity of V456 Cyg is negligible, and consistent with zero to within three times its very small error bars, so we assumed a circular orbit. Limb darkening (LD) was included using the quadratic law with theoretical coefficients taken from \citet{Claret17aa}. We fitted for the linear coefficient of each star and fixed the quadratic coefficients to the theoretical values. We also fitted low-order polynomials to the out-of-eclipse brightness of the system to remove any slow trends of brightness with time: in the case of V456\,Cyg it was found that a quadratic function for each of the four light curve segments (as plotted in Fig.\,\ref{fig:tess:lc}) was adequate. We obtained a good fit to the TESS data. 

To determine the uncertainties in the fitted parameters we ran the Monte Carlo and residual-permutation simulations implemented in {\sc jktebop} \citep{Me08mn}. Prior to this step we rescaled the errorbars in the SAP data to give a reduced $\chi^2$ of $\chi^2_\nu = 1$. The Monte Carlo simulations account for correlations between photometric parameters. The residual-permutation simulations are sensitive to low-frequency variability in data, so are useful to account for the effects of the pulsations present in this system. We adopted the larger of the two errorbars for each parameter, which in most cases was that from the Monte Carlo simulations. The third light comes out slightly negative but we have retained this in order to avoid underestimating the errorbars of the measured parameters -- a plausible reason for this is a slight overestimation of the sky background during the data reduction process. The resulting parameters and errorbars are collected in Table\,\ref{tab:jktebop}.

\begin{table} \centering
\caption{\label{tab:jktebop} Parameters of the V456\,Cyg system measured in Sect.\,\ref{sec:binarity}.}
\begin{tabular}{lc}    
\hline
\hline
Parameter                                       & Value                       \\
\hline                                                                        
$r_{\rm A}+r_{\rm B}$                           & $0.5228 \pm 0.0010$         \\ 
$k$                                             & $0.935 \pm 0.012$           \\ 
$i$ ($^\circ$)                                  & $83.198 \pm 0.052$          \\
$J$                                             & $0.7701 \pm 0.0063$         \\
$L_3$                                           & $-0.0344 \pm 0.0059$        \\
Linear LD coeff. for star A                     & $0.248 \pm 0.026$           \\
Linear LD coeff. for star B                     & $0.285 \pm 0.025$           \\
Quadratic LD coeff. for star A                  & 0.232 (fixed)               \\  
Quadratic LD coeff. for star B                  & 0.229 (fixed)               \\  
$P$ (d)                                         & $0.891193 \pm 0.000006$     \\  
$T_0$ (BJD/TDB)                                 & $2458709.73570 \pm 0.0000$  \\  
Semi-major axis $a$ (\Rsunnom)                  & $5.881 \pm 0.043$           \\
$K_{\rm A}$ (\kms)                              & $152.1 \pm 1.7$ $^*$        \\  
$K_{\rm B}$ (\kms)                              & $179.4 \pm 1.7$ $^*$        \\  
\hline
$r_{\rm A}$                                     & $0.2701 \pm 0.0015$         \\  
$r_{\rm B}$                                     & $0.2563 \pm 0.0019$         \\  
Light ratio                                     & $0.665 \pm 0.014$           \\  
Mass of star A (\Msunnom)                       & $1.859 \pm 0.041$           \\  
Mass of star B (\Msunnom)                       & $1.576 \pm 0.037$           \\  
Radius of star A (\Rsunnom)                     & $1.588 \pm 0.015$           \\  
Radius of star B (\Rsunnom)                     & $1.507 \pm 0.016$           \\
$\log g$ of star A (c.g.s.)                     & $4.306 \pm 0.006$           \\ 
$\log g$ of star B (c.g.s.)                     & $4.279 \pm 0.008$           \\ 
\Teff\ of star A (K)                            & $7750 \pm 100$ $^*$         \\ 
\Teff\ of star B (K)                            & $7250 \pm 120$              \\
\hline                              
\end{tabular}    
\tablefoot{Parameters with a superscripted N were calculated using the nominal physical constants and solar quantities defined by the IAU \citep{Prsa+16aj}. Starred quantities were taken from \citet{Bakis2014}. The upper part of the table contains fitted quantities and the lower part contains derived quantities.}
\end{table}                                                                                                                                              

\citet{Bakis2014} measured velocity amplitudes for the two stars from time-series spectroscopy, which we used to determine the full physical properties of the system (Table\,\ref{tab:jktebop}). This was done using the {\sc jktabsdim} code \citep{Me++05aa}, which implements standard equations and propagates errorbars using a perturbation analysis. We used the IAU nominal solar properties and physical constants \citep{Prsa+16aj} for compatibility with other work. Our mass measurements agree well with those found by \citet{Bakis2014}, as expected, but our radius measurements differ by roughly 5$\sigma$. Our results are likely more precise as they are based on much more extensive and high-precision TESS photometry; the discrepancy can be assigned to the pulsation signature present in the light curve. The masses of the stars are measured to 2.3\% precision: we are in the process of obtaining new spectroscopy to improve these measurements. Our surface brightness ratio indicates a secondary star effective temperature (\Teff) of 7250$\pm$120\,K, considerably higher than found by \citet{Bakis2014} (6755$\pm$400\,K), when we adopt their primary star \Teff\ value. 

We have determined the distance to V456\,Cyg using the physical properties of the system, apparent magnitudes in the $BV$ \citep{Hog+00aa} and $JHK_s$ bands \citep{Cutri+03book}, and the bolometric corrections from \citet{Girardi+02aa}. The requirement of consistent distances across the optical and infrared passbands could be satisfied by adopting a small reddening value of $E(B-V) = 0.05 \pm 0.01$\,mag, and returned distances in excellent agreement with the value of $587.5 \pm 4.3$\,pc from \textit{Gaia} EDR3 \citep{Gaia21aa}. This consistency supports the evidence of the reliability of the physical properties determined here and the \Teff\ measured by \citet{Bakis2014} for star~A.

\section{Asteroseismic analysis}
\label{sec:asteroseismology}
\subsection{Frequency analysis}
\label{subsec:freq-analysis}

After the binary modelling, we determined the pulsation frequencies by iteratively prewhitening the residual light curve (hereafter called the pulsation light curve), using the code from \citet{VanBeeck2021}. At each step, we selected the frequency $\nu_i$ with the highest $S/N$ in the Lomb-Scargle periodogram \citep{Scargle1982} of the light curve and determined the corresponding amplitude $A_i$ and phase $\phi_i$ by non-linearly fitting a sine wave of the form $A_i\sin\left(2\pi\nu_i \left(t-t_0\right) + \phi_i\right)$ to the light curve, using the mean timestamp of the photometric data points as the zero point $t_0$. Measured frequencies were selected until none remained with $S/N > 4$ \citep{Breger1993}, which was calculated within a $1\,\rm d^{-1}$ window in the Lomb-Scargle periodogram.

\begin{figure}
    \centering
    \includegraphics[width=88mm]{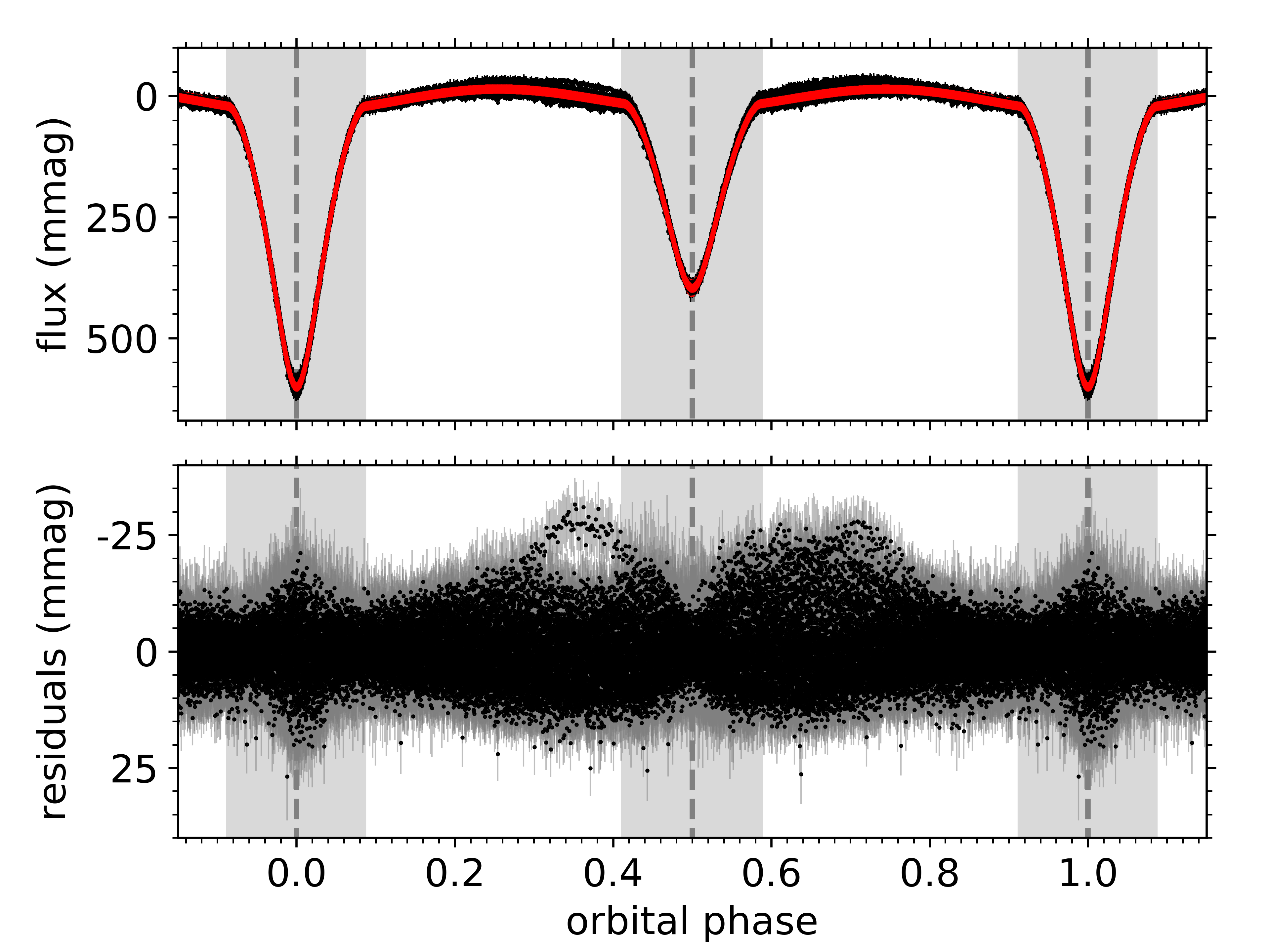}
    \caption{Reduced light curve of V456\,Cyg, phase-folded over the binary orbit. The eclipses are indicated in grey. \emph{Top:} original light curve (black) with the best-fitting binary model (red). \emph{Bottom:} residual light curve, called the pulsation light curve in this work.}
    \label{fig:V456Cyg_phasefold}
\end{figure}

\begin{figure*}
    \centering
    \includegraphics[width=\textwidth]{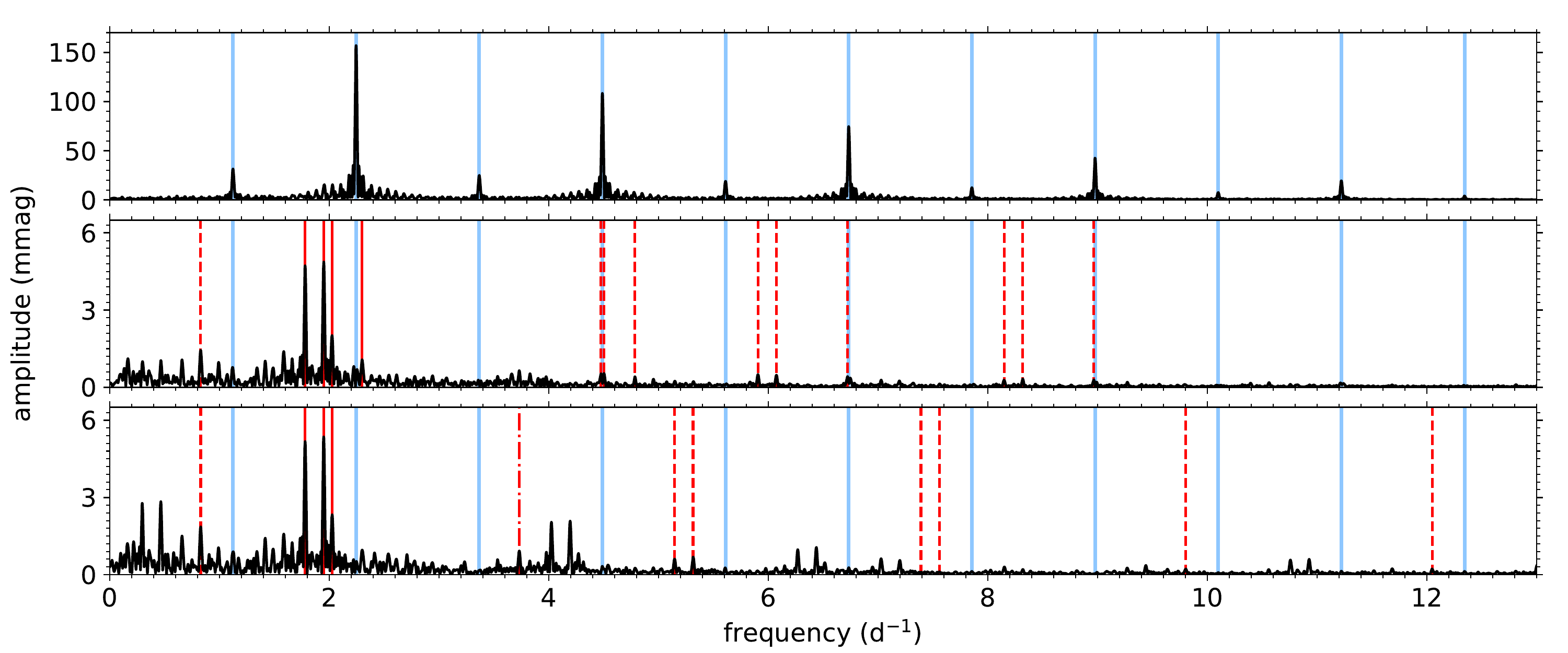}
    \caption{Lomb-Scargle periodograms of the reduced light curve of V456\,Cyg. {\em Top:} periodogram of the full reduced light curve of V456\,Cyg, with the harmonics of the orbital frequency $\nu_{\rm orb}$ indicated by full light-blue lines. {\em Middle:} periodogram of the full pulsation light curve. The frequencies that were detected from this light curve ($\nu_{a}$ to $\nu_{n}$ in Table\,\ref{tab:v456cyg_freq}) are marked in red. Full and dashed lines mark the independent frequencies and combination frequencies, respectively. {\em Bottom:} periodogram of the out-of-eclipse part of the pulsation light curve. The frequencies that were detected from this light curve ($\nu_{1}$ to $\nu_{11}$ in Table\,\ref{tab:v456cyg_freq}) are marked in red. Full and dashed lines mark the independent frequencies and combination frequencies, respectively. The dash-dotted line is the only combination frequency that does not include the orbital frequency $\nu_{\rm orb}$.}
    \label{fig:V456Cyg_scargle}
\end{figure*}

First, we prewhitened the full pulsation light curve, shown in Fig.\,\ref{fig:V456Cyg_phasefold}. The frequency values are listed in Table\,\ref{tab:v456cyg_freq} ($\nu_a$ to $\nu_n$) and illustrated in the middle of Fig.\,\ref{fig:V456Cyg_scargle}. Four of these fourteen frequencies are independent, with typical values for prograde dipole g-modes with $(\ell,m) = (1,1)$. The other frequencies are combinations, spaced integer multiples of $\nu_{\rm orb}$ away from one of the two dominant pulsation frequencies, or differ less than the frequency resolution $f_{\rm res} = 1/T$ from an orbital harmonic frequency, where $T$ is the time span covered by the light curve. This can be explained by the amplitude modulation of the light curve during the eclipses, as seen in Fig.\,\ref{fig:V456Cyg_phasefold}. The observed pulsation amplitudes are larger (smaller) during the primary (secondary) eclipse, indicating that the g~modes belong to the secondary component of V456\,Cyg.

We then compared these results with frequencies measured for the out-of-eclipse part of the pulsation light curve, shown in the white sections of Fig.\,\ref{fig:V456Cyg_phasefold}. These out-of-eclipse frequencies are listed in the bottom half of Table\,\ref{tab:v456cyg_freq} ($\nu_1$ to $\nu_{11}$) and are illustrated in the bottom panel of Fig.\,\ref{fig:V456Cyg_scargle}. As expected, most of the previously found $\nu_{\rm orb}$-spacings were caused by the amplitude modulations during the eclipses, and are no longer present. However, several new $\nu_{\rm orb}$-spacings and combinations are detected. This indicates further orbital-phase-dependent amplitude modulation of the observed pulsations in the out-of-eclipse part of the light curve, and points to possible tidal perturbation of these g-mode pulsations. This is discussed in detail in Sect.\,\ref{subsec:tidal-tilted}.

 \begin{table}[ht]
 \caption{\label{tab:v456cyg_freq} Overview of the measured frequencies, including the identified combinations (within a 3-$\sigma$ limit). \emph{Top:} frequencies obtained from the full pulsation light curve. Frequencies marked with $({}^*)$ differ less than the frequency resolution $f_{\rm res}$ from a harmonic of $\nu_{\rm orb}$. \emph{Bottom:} frequencies obtained from the out-of-eclipse pulsation light curve.}
\centering \small
\begin{tabular}{rrrrrl} 
\hline\hline
{} & frequency & amplitude & phase & S/N & combinations\\
{} & ($\rm d^{-1}$) & (mmag) & (rad) & & \\
\hline\vspace{-2.5mm}\\
$\nu_{a}$   & $0.8280(4)$ & $1.39(6)$ & $1.39(4)$  & $4.5$ & $\nu_c$ - $\nu_{\rm orb}$\\
$\nu_{b}$   & $1.7800(1)$ & $4.63(6)$ & $-0.64(1)$ & $9.4$ &  \\
$\nu_{c}$   & $1.9495(1)$ & $4.83(6)$ & $2.65(1)$  & $7.5$ &  \\
$\nu_{d}$   & $2.0257(2)$ & $1.98(6)$ & $1.22(3)$  & $6.3$ &  \\
$\nu_{e}$   & $2.2958(5)$ & $1.11(6)$ & $-0.21(5)$ & $5.1$ &  \\
$\nu_{f}^*$   & $4.474(1)$  & $0.47(6)$ & $-0.6(1)$  & $4.0$ &  4$\nu_{\rm orb}$ \\
$\nu_{g}^*$   & $4.504(1)$  & $0.42(6)$ & $2.8(1)$   & $4.7$ &  4$\nu_{\rm orb}$ \\
$\nu_{h}$   & $4.783(2)$  & $0.35(6)$ & $-0.5(2)$  & $4.0$ & 6$\nu_{\rm orb}$ - $\nu_c$\\
$\nu_{i}$   & $5.906(1)$  & $0.49(6)$ & $0.9(1)$   & $6.7$ & 7$\nu_{\rm orb}$ - $\nu_c$\\
$\nu_{j}$  & $6.073(1)$  & $0.47(6)$ & $-2.1(1)$  & $5.8$ & 7$\nu_{\rm orb}$ - $\nu_b$\\
$\nu_{k}^*$  & $6.721(1)$  & $0.41(6)$ & $2.4(1)$   & $5.5$ & 6$\nu_{\rm orb}$ \\
$\nu_{l}$  & $8.149(2)$  & $0.26(6)$ & $-2.6(2)$  & $5.6$ & 9$\nu_{\rm orb}$ - $\nu_c$\\
$\nu_{m}$  & $8.318(2)$  & $0.32(6)$ & $0.8(2)$   & $5.6$ & 9$\nu_{\rm orb}$ - $\nu_b$\\
$\nu_{n}^*$  & $8.965(2)$  & $0.27(6)$ & $-1.1(2)$  & $5.2$ & 8$\nu_{\rm orb}$ \\
\hline\vspace{-2.5mm}\\
$\nu_{1}$  & $0.8289(4)$ & $1.77(6)$ & $1.06(4)$  & $4.3$ & $\nu_3$ - $\nu_{\rm orb}$\\
$\nu_{2}$  & $1.7798(1)$ & $5.05(6)$ & $-1.38(1)$ & $9.0$ & \\
$\nu_{3}$  & $1.9491(1)$ & $5.29(6)$ & $1.89(1)$  & $7.3$ & \\
$\nu_{4}$  & $2.0265(2)$ & $2.27(6)$ & $0.38(3)$  & $6.2$ & \\
$\nu_{5}$  & $3.7292(8)$ & $0.86(6)$ & $2.21(7)$  & $4.5$ & $\nu_2 + \nu_3$\\
$\nu_{6}$  & $5.1463(9)$ & $0.76(6)$ & $-1.53(8)$ & $5.6$ & $\nu_2$ + $3\nu_{\rm orb}$\\
$\nu_{7}$  & $5.3142(7)$ & $0.99(6)$ & $2.17(6)$  & $5.6$ & $\nu_3$ + $3\nu_{\rm orb}$\\
$\nu_{8}$  & $7.390(2)$  & $0.36(6)$ & $0.5(2)$   & $4.6$ & $\nu_2$ + $5\nu_{\rm orb}$\\
$\nu_{9}$  & $7.5590(5)$ & $1.02(6)$ & $-2.16(6)$ & $5.2$ & $\nu_3$ + $5\nu_{\rm orb}$\\
$\nu_{10}$ & $9.8040(8)$ & $0.80(6)$ & $-0.15(8)$ & $4.7$ & $\nu_3$ + $7\nu_{\rm orb}$\\
$\nu_{11}$ & $12.049(1)$ & $0.44(6)$ & $1.9(1)$   & $4.1$ & $\nu_3$ + $9\nu_{\rm orb}$\\
\hline
\end{tabular} 
\end{table}

\subsection{Near-core rotation}
\label{subsec:core-rot}
\begin{figure}
    \centering
    \includegraphics[width=88mm]{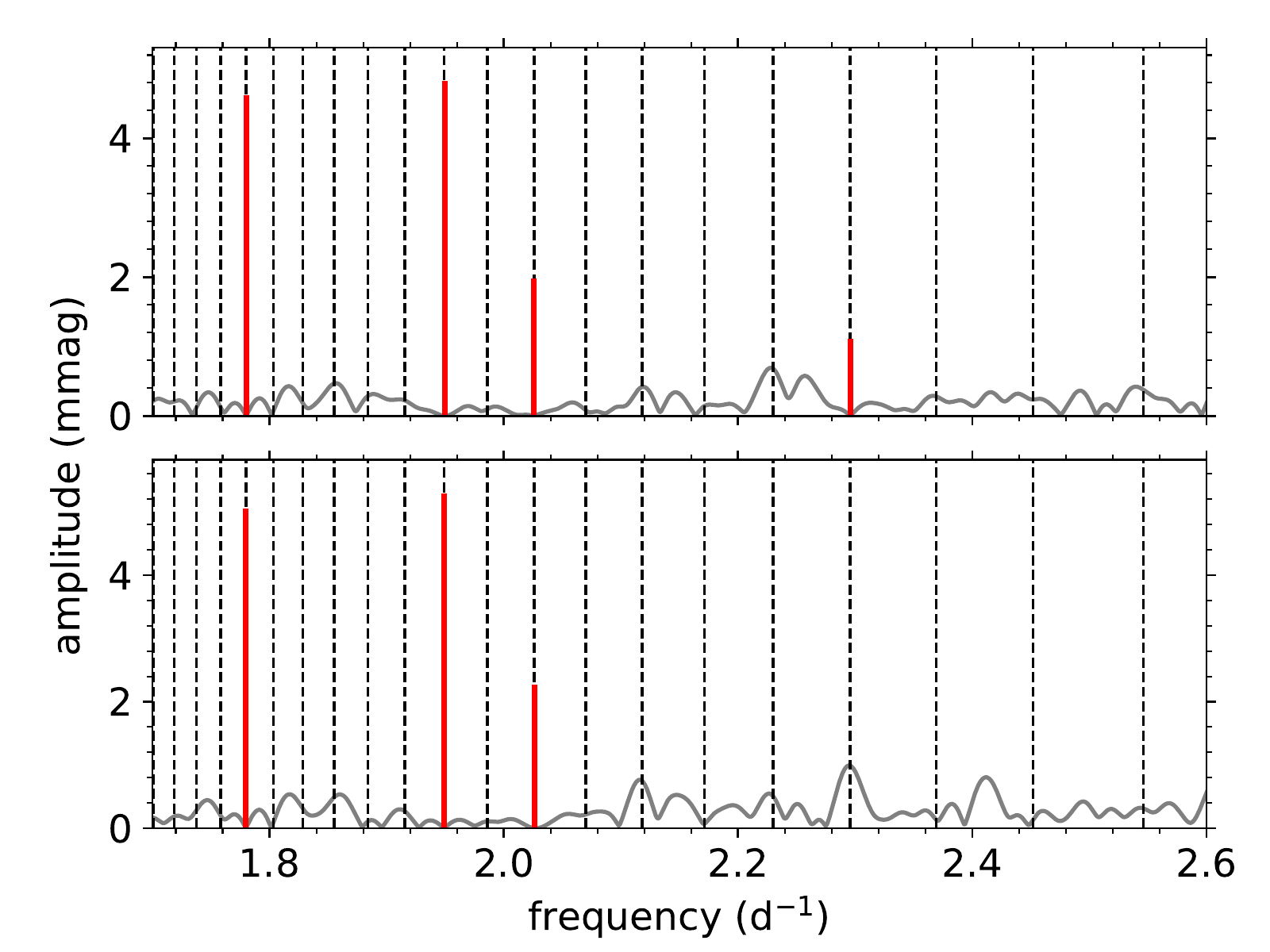}
    \caption{Best-fitting asymptotic g-mode pattern (dashed lines), with the measured independent pulsation frequencies (red full lines) and the Fourier transform of the residual light curve after prewhitening the measured pulsation frequencies. This is shown for both the full light pulsation curve ({\em top}) and the out-of-eclipse pulsation light curve ({\em bottom}).}
    \label{fig:g-mode-pattern}
\end{figure}

In the first part of our analysis, we determined the near-core rotation rate $\nu_{\rm rot}$ and the buoyancy travel time $\Pi_0$ of the secondary component by fitting the four independent g-mode frequencies between 1.75\,$\rm d^{-1}$ and 2.35\,$\rm d^{-1}$ ($\nu_b$ to $\nu_e$). Because g~modes in rotating stars are strongly influenced by the Coriolis force \citep[e.g.][]{Bouabid2013,Aerts2018} and less sensitive to non-spherical deformations of the star \citep{Henneco2021,Dhouib2021}, we ignored the tidal effects on the g-mode frequencies \citep[e.g.][]{Balona2018} and used the Traditional Approximation of Rotation \citep[TAR;][]{Eckart1960,Bildsten1996,Lee1997}, whereby the horizontal component of the rotation vector is neglected. Assuming that all four g~modes have the same geometry with respect to the rotation axis of the star, we fitted the frequencies $\nu_b$ to $\nu_e$ with asymptotic g-mode frequencies $$\nu_{n\ell m} = m\nu_{\rm rot} + \left(\frac{\Pi_0}{\sqrt{\lambda_{s\ell m}}}\left(n + \alpha_g\right)\right)^{-1},$$ where $(n,\ell,m)$ is the mode identification, $s = 2\nu_{\rm rot}/(\nu_{n\ell m}-m\nu_{\rm rot})$ the spin parameter, and $\alpha_g$ a phase term dependent on the boundaries of the g-mode cavity \citep{VanReeth2016}. While multiple valid solutions (listed in Table\,\ref{tab:rotation}) were found, the $\nu_{\rm rot}$~value of the first solution matches the orbital frequency $\nu_{\rm orb}$ well within its $99\%$-confidence interval. This indicates that the secondary component is synchronously rotating, as expected for such short-period binaries. Therefore, this is the most likely true solution. The corresponding theoretical g-mode pattern is illustrated in Fig.\,\ref{fig:g-mode-pattern}.

Additionally, two implicit assumptions were made in this analysis. First, the length of our light curve, 54 days, is insufficient to resolve most prograde dipole g-modes with consecutive radial orders (as can be seen in Fig.\,\ref{fig:g-mode-pattern}). Hence, we assumed that the amplitudes of the undetected g~modes are much smaller than those listed in Table\,\ref{tab:v456cyg_freq}, so that our measured pulsation frequencies were not affected. Second, we ignored possible modulations or glitches in the g-mode pattern \citep[e.g.][]{Miglio2008a,Ouazzani2020}, which can affect the measured values of $\nu_{\rm rot}$ and $\Pi_0$ \citep{Christophe2018}. Because of our small number of modelled frequencies, we likely overfitted the observations and underestimated the error margins of the model parameters. To compensate, we report the conservative error margins associated with the $99\%$-confidence interval, rather than the 1-$\sigma$ error margins. Hence, given our accurate derived $\nu_{\rm rot}$-value, we can conclude that these implicit conditions were also fulfilled.

 \begin{table}[ht]
 \caption{\label{tab:rotation} Rotation rates $\nu_{\rm rot}$ and buoyancy travel times $\Pi_0$ for the best-fitting asymptotic g-mode patterns, with conservative error margins derived from the $99\%$-confidence interval. The mode identifications $(n,\ell,m)$ are made with respect to the rotation axis. The $\nu_{\rm rot}$ value of the first pattern (in the top row) matches the orbital frequency $\nu_{\rm orb}$ of the binary within its error margins.}
\centering \small
\begin{tabular}{ccccccc} 
\hline\hline
\multicolumn{4}{c}{radial order $n$} & ($\ell$,$m$) & $\nu_{\rm rot}$ & $\Pi_0$\\
$\nu_b$ & $\nu_c$ & $\nu_d$ & $\nu_e$ &           & ($\rm d^{-1}$)  & (s)\\
\hline\vspace{-2.5mm}\\
 18 & 23 & 25 & 31 & (1,1) & 1.113\,(14) & 4380\,(140)\\
 20 & 27 & 30 & 40 & (1,1) & 1.2886\,(8) & 4505\,(10)\\
 24 & 33 & 37 & 51 & (1,1) & 1.3514\,(3) & 4076\,(6)\\
\hline
\end{tabular} 
\end{table}

\subsection{Evidence for tidally perturbed pulsations}
\label{subsec:tidal-tilted}
As discussed in Sect.\,\ref{subsec:freq-analysis}, the detected g-modes are part of $\nu_{\rm orb}$-spaced multiplets and exhibit orbital-phase-dependent amplitude modulation. There are multiple physical processes that can cause this, including the phase-dependent light contribution of the pulsating star in the binary \citep[e.g.][]{Steindl2021}, Doppler beaming \citep[e.g.][]{Zucker2007}, the light-travel-time effect \citep[e.g.][]{Murphy2015}, tidally excited \citep[e.g.][]{Fuller2017} and tidally perturbed pulsations \citep[e.g.][]{Reyniers2003b,Reyniers2003a,Fuller2020}. 

\begin{figure}
    \centering
    \includegraphics[width=88mm]{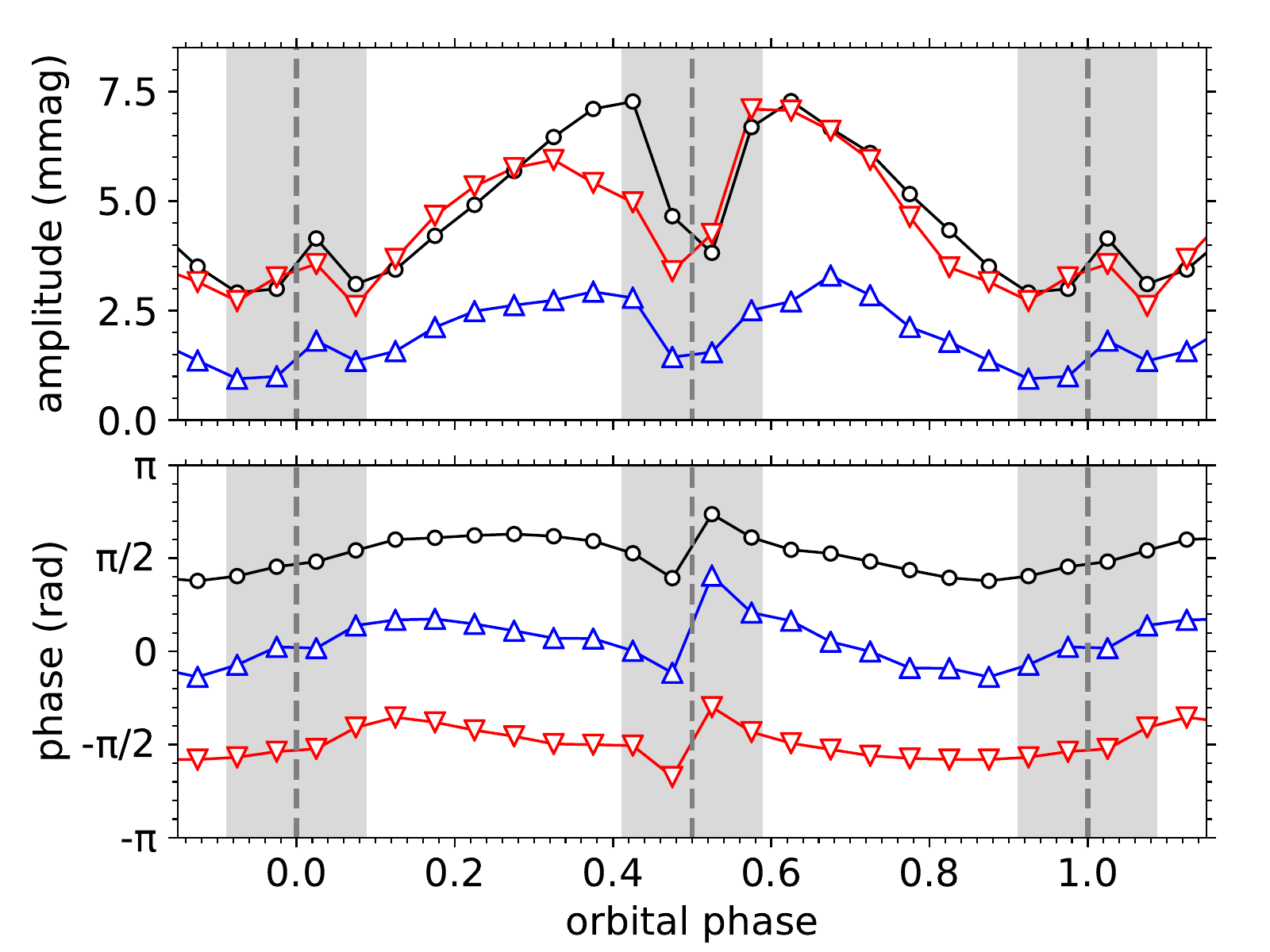}
    \caption{Amplitude ({\em top}) and phase ({\em bottom}) variability of the three dominant independent pulsations of V456\,Cyg, measured from the out-of-eclipse residual light curve: $\nu_2$ (red downward triangles), $\nu_3$ (black circles) and $\nu_4$ (blue upward triangles). The eclipses are indicated in grey, with a dashed line marking their centres.}
    \label{fig:V456Cyg_amp-phase-var}
\end{figure}

To determine the cause(s), we divided the observed data points in different bins as a function of orbital phase, and non-linearly fitted the three dominant g-modes ($\nu_b$ to $\nu_d$; $\nu_2$ to $\nu_4$) within each bin, keeping the frequencies fixed. The results were validated by doing the analysis for different numbers of bins (varying from 5 to 40), and are illustrated in Fig.\,\ref{fig:V456Cyg_amp-phase-var} for 20 bins (with $\sim$1800 data points per bin). We can see that all three g~modes behave similarly. The pulsation phases vary slowly and contemporaneously during the out-of-eclipse phases and exhibit jumps of $\pi/2$\,rad during the secondary eclipse. Their amplitudes are maximal just before and after the secondary eclipse, and appear to be larger (smaller) during the primary (secondary) eclipse because the light contribution by the pulsating star varies. We verified this by estimating the relative pulsation amplitude modulation at the secondary eclipse as $$\delta A = \frac{1 + L_{2/1}^{-1}}{1 + L_{2/1, {\rm ecl}}^{-1}},$$ where $L_{2/1}$ is the binary light ratio given in Table\,\ref{tab:jktebop} and $L_{2/1, {\rm ecl}}$ is the effective light ratio at mid-eclipse, calculated as $$L_{2/1, {\rm ecl}} = (1 + L_{2/1})10^{-0.4\Delta m_{\rm ecl}} - 1,$$ with $\Delta m_{\rm ecl}$ the secondary eclipse depth in magnitude. For V456\,Cyg we obtained $\delta A = 0.34(2)$, but because the secondary eclipse signal is smeared out in the orbital phase bins used in Fig.\,\ref{fig:V456Cyg_amp-phase-var}, the amplitude modulations shown in this Figure are less pronounced. 

A first possible explanation for the pulsation modulations, the geometry of the binary system, is insufficient. The orbital-phase-dependent behaviour of the different g-modes is similar, but not identical. The modulations would not differ between pulsations if they were caused by ellipsoidal variability, Doppler beaming or the light-travel-time effect. Furthermore, the amplitude and phase variations during the out-of-eclipse phases are too large to be explained by the binary geometry. The pulsation amplitudes vary by a factor three, while the phase variations are $\sim$200 times larger than expected from the binary orbit time delays.

Tidally excited pulsations are also not the cause of the observed modulations. As we have shown in Sect.\,\ref{sec:binarity} and Sect.\,\ref{subsec:core-rot}, V456\,Cyg is a synchronised and circularised binary. Hence, the tides cannot dynamically excite pulsation modes here \citep{Guo2021}. Additionally, neither the independent g-mode frequencies nor their combinations coincide with orbital harmonic frequencies, as we can see in Fig.\,\ref{fig:V456Cyg_scargle}.

Therefore, tidal perturbation of the pulsations is the best explanation for the pulsation amplitude and phase variations of V456\,Cyg. As discussed in Sect.\,\ref{subsec:core-rot}, the tidal perturbation of the g-mode pulsation cavity is sufficiently small that the impact on the g-mode frequencies appears to be negligible, but the observed modulations of the amplitudes and phases are much stronger.

The tidal perturbation of pulsations can also include tidal tilting. Indeed, if we account for the eclipse modulations of the pulsation amplitudes in Fig.\,\ref{fig:V456Cyg_amp-phase-var}, the observed amplitude variability of the g~modes is observationally reminiscent of the model developed by \citet{Fuller2020} for $(\tilde{\ell},\tilde{m}) = (1,0)$~modes in HD\,74423. But as shown by \citet{Fuller2020}, pure tilted $(\tilde{\ell},\tilde{m}) = (1,0)$-modes are expected to exhibit either 0\,rad or $\pi$\,rad pulsation phase modulations over the binary orbit. Our observed pulsation phase modulations are small but non-zero, which indicates that the pulsations have non-axisymmetric components \citep[with respect to the tidal axis;][]{Fuller2020}, consistent with our pulsation mode identification with respect to the rotation axis, $(\ell,m) = (1,1)$. However, because the current theoretical framework for tidally tilted pulsations does not account for the Coriolis force, a more detailed evaluation of the true pulsation axis lies outside the scope of this work. The pulsations of V456\,Cyg are high-order g~modes in the subinertial regime, where the contribution of the Coriolis force is non-negligible and has to be treated non-perturbatively \citep{Aerts2021}.

\begin{figure}
    \centering
    \includegraphics[width=88mm]{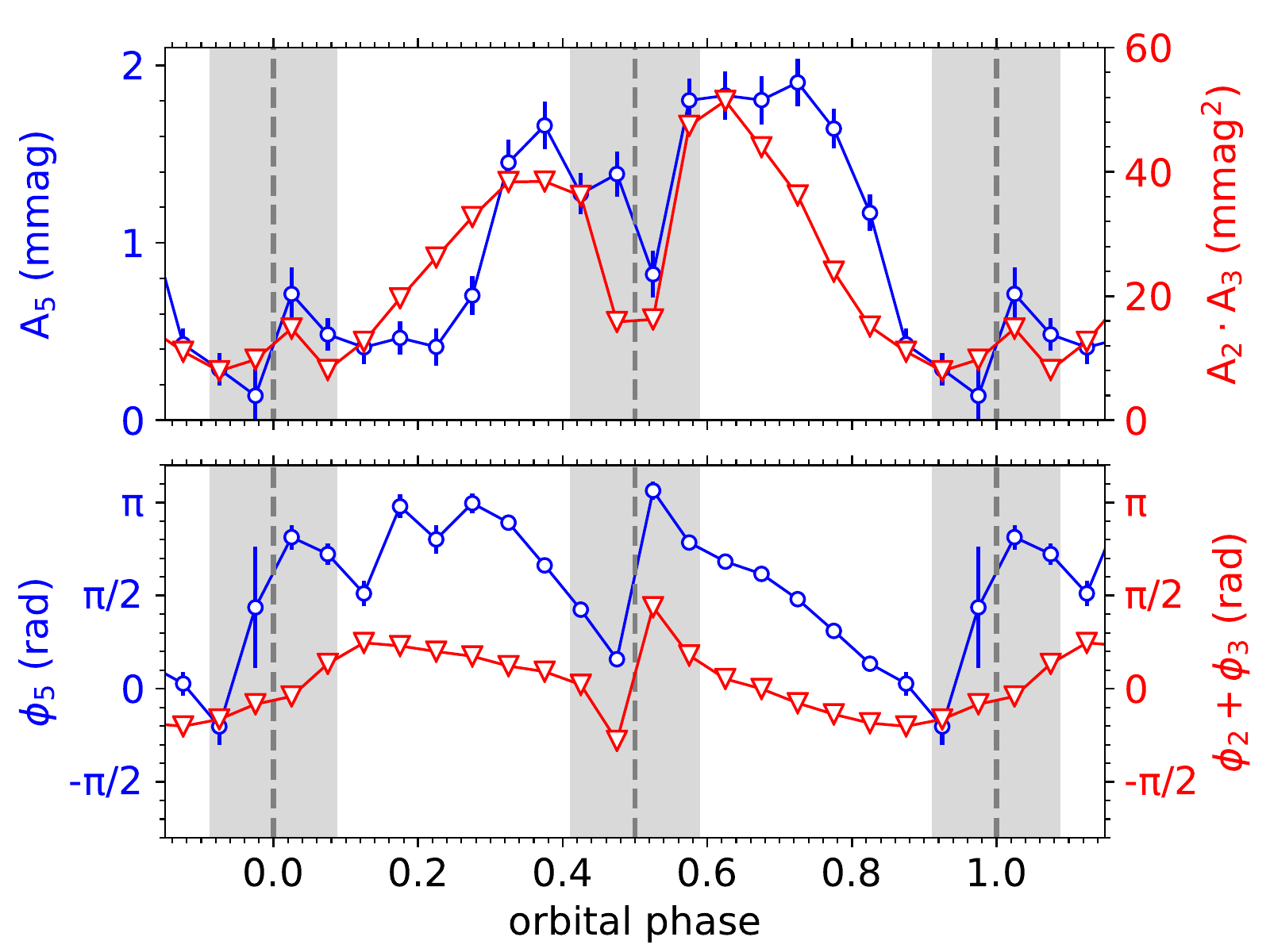}
    \caption{Amplitude ({\em top}) and phase ({\em bottom}) variability of the combination frequency $\nu_5 = \nu_2 + \nu_3$ (blue circles), measured from the out-of-eclipse residual light curve, as a function of the orbital phase. These are compared with the relative amplitude scaling factor $A_2\cdot A_3$ and the sum of the parent frequency phases $\phi_2 + \phi_3$ \citep[red downward triangles; e.g.][]{Buchler1997,VuilleBrassard2000}.}
    \label{fig:V456Cyg_comb-freq}
\end{figure}

Additionally, tides are known to affect nonlinear pulsation mode coupling \citep[e.g.][]{Fuller2020,Guo2021}. To further evaluate the influence of tides on the g-mode coupling in V456\,Cyg, we analysed the coupled modes $\nu_2$, $\nu_3$ and $\nu_5$. Based on the values listed in Table\,\ref{tab:v456cyg_freq}, these pulsations fulfil the formal criteria of coupled modes, $\sum_i\nu_{\rm parent,i} = \nu_{\rm child}$ and $\sum_i\phi_{\rm parent,i} = \phi_{\rm child} + k\pi/2$ with $k \in \mathbb{Z}$ \citep[e.g.][]{Buchler1997,VuilleBrassard2000}, within 2$\sigma$. This is investigated further in Fig.\,\ref{fig:V456Cyg_comb-freq}, where we show the amplitude and phase variability of $\nu_5$ as a function of the orbital phase, and compare them with the product of the amplitudes and the sum of the phases of $\nu_2$ and $\nu_3$, respectively. These quantities are commonly used to investigate the non-linear coupling and the relative damping and excitation rates of the pulsations \citep[e.g.][]{Dziembowski1982,Lee2012,Saio2018}. For example, \citet{Bowman2016} used them to define the coupling factor $\mu$ between parent frequencies $\nu_{\rm p1}$ and $\nu_{\rm p2}$ and the child frequency $\nu_{\rm c}$ as $\mu = A_{\rm c}/(A_{\rm p1}A_{\rm p2})$, to investigate coupling between p-mode pulsations. Here in our work, we see in Fig.\,\ref{fig:V456Cyg_comb-freq} that there is a small orbital phase shift between the amplitude variability of $\nu_5$ and of $\nu_2$ and $\nu_3$. Moreover, the phases $\phi_5$ and $\phi_2+\phi_3$ differ as a function of the orbital phase, so the pulsation phase criterion of coupled modes is no longer fulfilled. When the amplitude of $\nu_5$ is dominant during the out-of-eclipse orbital phases, the difference between the phases $\phi_5$ and $\phi_2+\phi_3$ steadily decreases. These differences may provide detailed information about the role of tides in g-mode coupling in binaries.

\section{Discussion and conclusions}
\label{sec:conclusions}
We report the second detection of tidally perturbed g-mode pulsations in the short-period binary V456\,Cyg, after the detection in $\pi^5$~Ori by \citet{Jerzykiewicz2020}. Other reports in the literature of tidally perturbed or tilted pulsations were limited to p~modes and mixed p- and g-mode pulsations. 

The observed g~modes belong to the secondary component, which has a uniform radial rotation profile, with $\nu_{\rm rot} = 1.113(14)\,\rm d^{-1}$, synchronous with the binary orbital frequency. The measured value of the buoyancy travel time $\Pi_0$, $4377\pm140$s, places the secondary well within the $\gamma$\,Dor instability strip, in agreement with the derived values for its mass ($1.576\pm0.037\rm M_\odot$) and $T_{\rm eff}$ ($7250\pm120$K).

The observed g~mode pulsations exhibit orbital-phase-dependent amplitude and phase variability. The three most dominant g~modes all show similar amplitude variability, reminiscent of the theoretical models developed by \citet{Fuller2020} for HD\,74423. Their amplitudes are amplified around the secondary eclipse and modulated by the reduced flux from the secondary component during the eclipse itself. This indicates that the g~modes are amplified on the side of the secondary facing the primary in V456\,Cyg. By contrast, the g-mode frequencies are much less affected by the tides, as demonstrated by our successful asymptotic g-mode modelling using the TAR, which assumes spherical symmetry. There is also an indication of tidal influence on the non-linear g-mode coupling in V456\,Cyg, given by the comparison between the amplitudes and phases of the frequencies $\nu_2$, $\nu_3$ and $\nu_5$ ($=\nu_2+\nu_3$). The formal criteria to classify these as coupled pulsation modes \citep{VuilleBrassard2000} are fulfilled when they are evaluated using the entire out-of-eclipse pulsation light curve. However, this is not the case when these pulsations are evaluated as a function of the orbital phase, providing strong observational indications of tidal modulation of mode coupling in V456\,Cyg.

The discovery of tidally perturbed g~modes in V456\,Cyg presents us with a new opportunity. Since g-mode pulsators have dense pulsation frequency spectra, such targets require long time series of space-based data to resolve individual pulsations \citep[e.g.][]{Li2020,Garcia2022}. However, the radial orders $n$ of the observed g-mode pulsations of V456\,Cyg are a bit lower than usual \citep{Li2020}, with values between 18 and 31. This part of the g-mode spectrum is less dense, and because the dominant pulsation modes have much higher amplitudes than the neighbouring modes with consecutive radial orders, we were able to resolve these dominant g~modes, despite only having 54 days of observations. Moreover, previous detections of tidally perturbed or tilted pulsations were limited to p-mode and mixed p- and g-mode pulsators. Thanks to the existing theoretical frameworks \citep[e.g.][]{Miglio2008a,Bouabid2013,Mathis2009,Prat2019,Ouazzani2020}, g-mode pulsations are often easier to analyse than p~modes in moderate- to fast-rotating stars, making V456\,Cyg an interesting target for further follow-up studies. We expect that in future, more detections of tidally perturbed g-mode pulsations will be reported in the literature.

\begin{acknowledgements}
TVR gratefully acknowledges a postdoctoral fellowship from the Research Foundation Flanders (FWO) with grant agreement N\textdegree 12ZB620N. JVB acknowledges receiving support from the Research Foundation Flanders (FWO) under grant agreement N\textdegree V421221N.  DMB gratefully acknowledges a senior post-doctoral fellowship from the Research Foundation Flanders (FWO) with grant agreement N\textdegree 1286521N. The research leading to these results received partial funding from the KU Leuven Research Council (grant C16/18/005: PARADISE). We are grateful to Jim Fuller for useful discussions and we thank the referee, Don Kurtz, for his useful and constructive comments which improved the contents of this paper.

This paper includes data collected with the TESS mission, obtained from the MAST data archive at the Space Telescope Science Institute (STScI). Funding for the TESS mission is provided by NASA's Science Mission Directorate. We thank the whole team for the development and operations of the mission. STScI is operated by the Association of Universities for Research in Astronomy, Inc., under NASA contract NAS 5-26555. This research made use of the SIMBAD database, operated at CDS, Strasbourg, France, NASA’s Astrophysics Data System Bibliographic Services, and the VizieR catalogue access tool, CDS, Strasbourg, France.

This research also made use of Astropy\footnote{http://www.astropy.org} \citep[a community-developed core Python package for Astronomy;][]{astropy:2013, astropy:2018}, lmfit \citep{lmfit100}, Matplotlib \citep[the Python library for publication quality graphics;][]{Hunter2007}, and Numpy \citep{numpy}.
\end{acknowledgements}

\bibliographystyle{aa}
\bibliography{V456Cyg}

\end{document}